 \documentclass{emulateapj}

\newcommand{\spitzer}{{\it Spitzer}}
\newcommand{\herschel}{{\it Herschel}}
\newcommand{\nustar}{{\it NuSTAR}}
\newcommand{\swift}{{\it Swift}}
\newcommand{\xmm}{{\tt XMM-Newton}}
\newcommand{\suzaku}{{\it Suzaku}}
\newcommand{\chandra}{{\it Chandra}}
\newcommand{\swiftxrt}{{\it Swift}/XRT}
\newcommand{\swiftbat}{{\it Swift}/BAT}
\newcommand{\bepposax}{{\it BeppoSAX}}
\newcommand{\xspec}{{\tt Xspec}}
\newcommand{\mekal}{{\tt mekal}}
\newcommand{\pexrav}{{\tt pexrav}}

\newcommand{\mytorus}{{\tt MYtorus}}
\newcommand{\bntorus}{{\tt BNtorus}}
\newcommand{\iras}{IRAS\,09104$+$4109}

\usepackage{hyperref}
\usepackage{breakurl}

\slugcomment{}

\shorttitle{X-ray/IR study of a dusty hyperluminous QSO}
\shortauthors{Farrah et al.}

\begin{document}

\title{The Geometry of the Infrared and X-ray Obscurer in a Dusty Hyperluminous Quasar}
\author{Duncan Farrah\altaffilmark{1}}
\author{Mislav Balokovi{\'c}\altaffilmark{2}}
\author{Daniel Stern\altaffilmark{3}}
\author{Kathryn Harris\altaffilmark{1,4}}
\author{Michelle Kunimoto\altaffilmark{5}}
\author{Dominic J. Walton\altaffilmark{3}}
\author{David M. Alexander\altaffilmark{6}}
\author{Patricia Ar{\'e}valo\altaffilmark{7}}
\author{David R. Ballantyne\altaffilmark{8}}
\author{Franz E. Bauer\altaffilmark{9,10,11,12}}
\author{Steven Boggs\altaffilmark{13}}
\author{William N. Brandt\altaffilmark{14,15,16}}
\author{Murray Brightman\altaffilmark{2}}
\author{Finn Christensen\altaffilmark{17}}
\author{David L. Clements\altaffilmark{5}}
\author{William Craig\altaffilmark{12,18}}
\author{Andrew Fabian\altaffilmark{19}}
\author{Charles Hailey\altaffilmark{13}}
\author{Fiona Harrison\altaffilmark{2}}
\author{Michael Koss\altaffilmark{21}}
\author{George B. Lansbury\altaffilmark{6}}
\author{Bin Luo\altaffilmark{22,23}}
\author{Jennie Paine\altaffilmark{1}}
\author{Sara Petty\altaffilmark{24}}
\author{Kate Pitchford\altaffilmark{1}}
\author{Claudio Ricci\altaffilmark{9,12}}
\author{William Zhang\altaffilmark{25}}

\altaffiltext{1}{Department of Physics, Virginia Tech, Blacksburg, VA  24061, USA} 
\altaffiltext{2}{Cahill Center for Astronomy and Astrophysics, California Institute of Technology, Pasadena, CA 91125, USA} 
\altaffiltext{3}{Jet Propulsion Laboratory, California Institute of Technology, Pasadena, CA 91109, USA} 
\altaffiltext{4}{Instituto de Astrofisica de Canarias (IAC), E-38200 La Laguna, Tenerife, Spain; Departamento de Astrofisica, Universidad de La Laguna (ULL), E-38205 La Laguna, Tenerife, Spain}  
\altaffiltext{5}{Astrophysics Group, Imperial College London, Blackett Laboratory, Prince Consort Road, London SW7 2AZ, UK}
\altaffiltext{6}{Department of Physics, Durham University, Durham DH1 3LE, UK}
\altaffiltext{7}{Instituto de F\'isica y Astronom\'ia, Facultad de Ciencias, Universidad de Valpara\'iso, Gran Bretana N 1111, Playa Ancha, Valpara\'iso, Chile}
\altaffiltext{8}{Center for Relativistic Astrophysics, School of Physics, Georgia Institute of Technology, 837 State Street, Atlanta, GA 30332-0430, USA}
\altaffiltext{9}{Instituto de Astrof\'{\i}sica, Facultad de F\'{i}sica, Pontificia Universidad Cat\'{o}lica de Chile, Casilla 306, Santiago 22, Chile} 
\altaffiltext{10}{Millennium Institute of Astrophysics, MAS, Nuncio Monse\~{n}or S\'{o}tero Sanz 100, Providencia, Santiago de Chile} 
\altaffiltext{11}{Space Science Institute, 4750 Walnut Street, Suite 205, Boulder, Colorado 80301} 
\altaffiltext{12}{EMBIGGEN Anillo, Concepci\'{o}n, Chile}
\altaffiltext{13}{Space Science Laboratory, University of California, Berkeley, CA 94720, USA}
\altaffiltext{14}{Department of Astronomy \& Astrophysics, 525 Davey Lab, The Pennsylvania State University, University Park, PA 16802, USA}
\altaffiltext{15}{Institute for Gravitation and the Cosmos, The Pennsylvania State University, University Park, PA 16802, USA}
\altaffiltext{16}{Department of Physics, 104 Davey Lab, The Pennsylvania State University, University Park, PA 16802, USA}
\altaffiltext{17}{DTU Space, National Space Institute, Technical University of Denmark, Elektrovej 327, DK - 2800 Lyngby, Denmark}
\altaffiltext{18}{Lawrence Livermore National Laboratory, Livermore, CA 94550, USA}
\altaffiltext{19}{Institute of Astronomy, Madingley Road, Cambridge, CB3 0HA, UK}
\altaffiltext{20}{Physics Department, Columbia University, New York, NY 10027, USA}
\altaffiltext{21}{Institute for Astronomy, Department of Physics, ETH Zurich, Wolfgang-Pauli-Strasse 27, CH-8093 Zurich, Switzerland}
\altaffiltext{22}{School of Astronomy and Space Science, Nanjing University, Nanjing, 210093, China}
\altaffiltext{23}{Key laboratory of Modern Astronomy and Astrophysics (Nanjing University), Ministry of Education, Nanjing 210093, China}
\altaffiltext{24}{Green Science Policy Institute, Berkeley, CA 94709, USA}
\altaffiltext{25}{NASA Goddard Space Flight Center, Greenbelt, MD 20771, USA}

\begin{abstract}
We study the geometry of the AGN obscurer in \iras, an IR-luminous, radio-intermediate FR-I source at $z$=0.442, using infrared data from \spitzer\ and \herschel, X-ray data from \nustar, \swift, \suzaku, and \chandra, and an optical spectrum from Palomar. The infrared data imply a total rest-frame 1-1000$\mu$m luminosity of $5.5\times10^{46}$~erg\,s$^{-1}$ and require both an AGN torus and starburst model. The AGN torus has an anisotropy-corrected IR luminosity of $4.9\times10^{46}$~erg\,s$^{-1}$, and a viewing angle and half opening angle both of approximately $36\degr$ from pole-on.
The starburst has a star formation rate of $(110\pm34)$\,M$_{\odot}$ yr$^{-1}$ and an age of $<50$\,Myr. These results are consistent with two epochs of luminous activity in \iras: one approximately $150$\,Myr ago, and one ongoing. The X-ray data suggest a photon index of $\Gamma \simeq 1.8$ and a line-of-sight column of $N_{\rm H} \simeq 5\times10^{23}$~cm$^{-2}$. This argues against a reflection-dominated hard X-ray spectrum, which would have implied a much higher $N_{\rm H}$ and luminosity. The X-ray and infrared data are consistent with a bolometric AGN luminosity of $L_{\rm bol}\sim(0.5-2.5)\times10^{47}$\,erg\,s$^{-1}$. The X-ray and infrared data are further consistent with coaligned AGN obscurers in which the line of sight `skims' the torus. This is also consistent with the optical spectra, which show both coronal iron lines and broad lines in polarized but not direct light. Combining constraints from the X-ray, optical, and infrared data suggests that the AGN obscurer is within a vertical height of $20$\,pc, and a radius of 125\,pc, of the nucleus. 
\end{abstract}


\section{Introduction}
A significant fraction of galaxy assembly at $z\gtrsim0.5$ proceeds via episodes of rapid star formation (hundreds to thousands of Solar masses per year) and accretion onto supermassive black holes at a non-negligible fraction of the Eddington limit \citep[e.g.][]{Lilly1996,Dickinson2003,PerezGonzalez2005,far08,Wuyts2011,Bethermin2012,ale12,Madau2014,mrr16}. Moreover, there is evidence for a deep connection between starburst and AGN activity at all redshifts, from, for example, the $\rm{M}_{bh}-\sigma$ relation \citep[e.g.][]{Magor98,Tremaine2002}, and from the presence of starbursts and AGN in the same systems \citep{gen98,farrah03,alexander05,lon06,hernan09,harris16} at rates much higher than expected by chance. There is also evidence that star formation and AGN activity can directly affect each other (see \citealt{fabian12} for a review), via both quenching \citep[e.g.][]{croton06,chung11,farrah12,sch15,alat15} and triggering \citep[e.g.][]{king05,ish12,gaib12,silk13,Zubovas2013}.

The connection between star formation and AGN activity is challenging to study, for two reasons. First, the bulk of these activities occur at high redshifts, $1 \lesssim z \lesssim 7$ \citep[e.g.][]{chapman05,Richards2006,wang13,Delvecchio2014}, where they are seen both faintly and at coarsened spatial scales. Second, star forming regions and AGN are often occulted by large column densities of gas and dust. Thus, a substantial fraction of their light is observed in the infrared \citep{lag05,alexander05,igles07,farrah13,burg13,Bridge2013,mig13,cas14,vig14,lan15,gru16}. A choate picture of how star formation and AGN activity contribute to galaxy assembly thus requires both deep and wide blank-field extragalactic surveys, and case studies of individual objects at lower redshifts. The latter serve to create archetypes at high sensitivity and spatial resolution for how star formation and AGN activity proceed in galaxies, and to illustrate how constraints from multi-wavelength data can be combined. 

\iras \ \citep{klei88} at $z=0.442$ \citep{hew10} is one such archetype, for the relationship between luminous, obscured AGN and star formation. In the radio it is a `radio-intermediate' FR-I source, with a linear core and double-lobed structure \citep{hines93,osullivan+2012}. It is extremely IR-luminous \citep{mrr00,ruiz10,vignali+2011} with a rest-frame 1-1000\,$\mu$m luminosity of $\sim4\times10^{46}$\,erg\,s$^{-1}$, of which at least 70\% arises from AGN activity. The mass of free baryons in the system is however small compared to other IR-luminous systems, with only $\sim3.2\times10^9$\,M$_{\odot}$ of molecular Hydrogen and of order $10^7$\,M$_{\odot}$ of warm dust \citep{evans98,comb11}. Its optical spectrum is that of a Sy2 \citep{klei88,soi96,ver06}, but with broad H$\beta$, H$\gamma$ and \ion{Mg}{2} lines in polarized light \citep{hines93,tran00}. There is also a polarized, bipolar reflection nebula centered on the nucleus \citep{hines99}. Its optical spectrum further reinforces its extreme nature; for example its $[$\ion{O}{3}$]\lambda5007$\,\AA\ luminosity, at $7.7\times10^{43}$\,erg\,s$^{-1}$, is nearly an order of magnitude higher than any other type 2 quasar at $z<0.5$ \citep{lansbury+2015}. Inferring a current star formation rate from the equivalent width (EW) of [\ion{O}{2}]$\lambda3727\,$\AA\, yields $41\pm12$\,M$_{\odot}$yr$^{-1}$ \citep{bild08}. There is also evidence, from fitting model star formation histories to UV through optical photometry, for an episode of star formation 100--200\,Myr ago \citep{pip09}. Optical imaging and integral field spectroscopy reveal a disturbed system with several bright `knots' within its stellar envelope, of which one may be a second nucleus, multiple companions within 100\,kpc, and extended, [\ion{O}{3}] bright filaments \citep{soi96,craw96,armus99}.

\iras\ is a cD galaxy within the rich cluster MACS\,J0913.7$+$4056 \citep{klei88,hall97,far04}. This cluster is associated with spatially extended X-ray emission with a strong cool core \citep{fabian95,craw96}. Other examples of cool-core clusters hosting powerful AGN at $z<1$ are known, including H\,1821$+$643 \citep{russell10} and the Phoenix cluster \citep{mcdonald15}. Two cavities are visible in the X-ray emission, coincident with the radio hotspots \citep{hla12}. \iras\ itself is luminous in the X-ray \citep{fabian94}. The soft X-ray emission is dominated by plasma with a temperature of $\sim$5\,keV \citep[e.g.][]{franceschini+2000,osullivan+2012}. A hard component starts to contribute above $5$\,keV and dominates above $8\,$~keV. Two origins have been proposed for the hard component: the intrinsic AGN continuum transmitted along a line of sight absorbed by a column density of $\sim5\times10^{23}\,$~cm$^{-2}$, or reflection from cold material surrounding the X-ray source. The latter possibility requires a Compton-thick column density ($\gtrsim5\times10^{24}\,$~cm$^{-2}$) along the line of sight in order to completely obscure the intrinsic continuum. The X-ray-based determination of the intrinsic luminosity depends on which of these two scenarios dominates \citep[e.g.][]{franceschini+2000,iwasawa+2001,piconcelli+2007,vignali+2011,chiang+2013,lama14}.

Other than being an example of a key phase in AGN evolution, \iras\ is an excellent candidate for being the most luminous Compton-thick quasar at $z\lesssim0.5$. It may thus be one of the few Compton thick objects that is bright enough for probing the obscurer structure at multiple wavelengths, from the infrared (IR) through X-ray. A larger sample of luminous obscured quasars at $0.1<z<0.5$ (all of which are at least a factor of $\sim5$ less luminous than \iras) has been studied with \nustar\ by \citet{lansbury+2014,lansbury+2015}, in addition to single-object studies at lower ($z=0.051$; \citealt{gandhi+2014}), as well as higher redshift ($z\approx2$; \citealt{delmoro+2014}). All targets in the \nustar\ survey of type~2 quasars have been found to have either Compton-thick obscuration, or high obscuration with column densities in the $10^{23}-10^{24}$~cm$^{-2}$ range. While short \nustar\ observations typically yield only weak detections of these sources, several have sufficient photon statistics for modeling the obscurer in detail. Together with \iras, they form a small but important sample of high-luminosity AGN that bridge the gap between well studied AGN in the local Universe (e.g., \citealt{brightman+2015}) and their counterparts at high redshift (e.g., \citealt{iwas05,stern+2014}).

This system has thus been the subject of several multi-wavelength studies \citep[e.g.][]{vignali+2011}. In this paper we combine a new X-ray observation from \nustar\ and a new optical spectrum from Palomar with all available archival X-ray and IR data to study both the geometry of the AGN obscurer, and current star formation, in \iras. We constrain the viewing angle, torus opening angle, and other geometric properties of the IR and X-ray emitting AGN obscurer, and clearly detect ongoing star formation in the host galaxy. We adopt a position for \iras \ of 09h13m45.49s, +40d56m28.22s (J2000) and assume $H_{0}$ = 70\,km\,s$^{-1}$\,Mpc$^{-1}$, $\Omega = 1$, $\Omega_{\Lambda} = 0.7$. We quote all luminosities in units of erg\,s$^{-1}$.

\section{Observations}

\subsection{Infrared \& Optical}\label{iroptobs}
We assembled IR data from several sources. Photometry at 3.6 and 5.8$\mu$m from the Infrared Array Camera (IRAC, \citealt{faz04}) on-board {\itshape Spitzer} \citep{wer04} were obtained from \citet{ruiz10}, and checked against the {\itshape WISE} public catalogues \citep{wri10,cut13}. A spectrum from the Infrared Spectrograph (IRS, \citealt{houck04}) on {\itshape Spitzer}, spanning observed-frame 6-34$\mu$m, was acquired from version LR6 of the Cornell Atlas of {\itshape Spitzer}/IRS Sources (CASSIS, \citealt{leb11}). The spectrum (AOR key 6619136) was taken in cycle 3 of {\itshape Spitzer} operations. The calibration of these data was checked against published {\itshape Spitzer} IRAC and MIPS data \citep{ruiz10}, and against {\itshape WISE}. Photometry at $70\mu$m, $100\mu$m and $160\mu$m were obtained from archival observations by the Photodetector Array Camera and Spectrometer (PACS, \citealt{pog10}) on-board {\itshape Herschel} \citep{pilb10}. The raw data were reduced with version 14 of the {\itshape Herschel} Interactive Processing Environment (HIPE, \citealt{ott10}), and flux densities were extracted using aperture photometry within HIPE. The $70\mu$m and $100\mu$m data were checked for consistency against the 60$\mu$m and $100\mu$m data  from the {\itshape Infrared Astronomical Satellite} ({\itshape IRAS}, \citealt{neu84}), both from \citet{wang10} and manual reprocessing of the {\itshape IRAS} data using the Scan Processing and Integration tool (SCANPI). Photometry at $250\mu$m, $350\mu$m and $500\mu$m were obtained from archival observations by the Spectral and Photometeric Imaging REceiver instrument (SPIRE; \citealt{Griffin2010}) onboard {\itshape Herschel}, and processed within HIPE. Finally, an 850$\mu$m flux density was obtained from \citet{deane01}. The photometry flux densities are presented in Table~\ref{irfluxes}. The IRS spectrum is presented in \citet{ruiz13} and in the SED plot, where it is plotted as multiple photometric points. 

Some IR data are not included in this compilation. We do not include data from {\itshape WISE} or {\itshape IRAS} since the {\itshape Spitzer} and {\itshape Herschel} data cover their wavelengths at higher sensitivity and improved spectral resolution. We also do not include data at wavelengths shortward of 3.6$\mu$m. Our aim is to constrain the properties of the obscured AGN (in particular the geometry of the obscurer), and any ongoing star formation (see \S\ref{iroptanalysis}). The integrated emission from older stars is almost certainly negligible at observed-frame wavelengths of 3.6$\mu$m and longer, but may contribute significantly at shorter wavelengths.

\begin{deluxetable}{lcc}
\tabletypesize{\scriptsize}
\tablecolumns{3}
\tablewidth{0pc}
\tablecaption{Assembled infrared photometry of \iras\ }
\tablehead{\colhead{Facility}&\colhead{Wavelength}&\colhead{Flux density}\\
\colhead{}&\colhead{$\mu$m}&\colhead{mJy}}
\startdata
Spitzer-IRAC   & 3.6 & $4.74\pm1.21$ \\
Spitzer-IRAC   & 5.8 & $26.4\pm7.11$ \\
Herschel-PACS  & 70  & $439\pm24$    \\
Herschel-PACS  & 100 & $319\pm18$    \\
Herschel-PACS  & 160 & $160\pm23$    \\
Herschel-SPIRE & 250 & $72\pm14$     \\
Herschel-SPIRE & 350 & $<50$       \\
Herschel-SPIRE & 500 & $<50$       \\
JCMT-SCUBA     & 850 & $<10$       \\
\enddata  
\tablecomments{The IR data also include the {\itshape Spitzer}-IRS spectrum in Figure \ref{fig:irsed}. The PACS flux density errors include uncertainties arising from celestial standard models \citep{balog14}. Upper limits are quoted at $3\sigma$ significance.}
\label{irfluxes}
\end{deluxetable}

We obtained an optical spectrum of \iras\ on UT 2014 December 23 using the Double Spectrograph (DBSP, \citealt{oke82}), a dual-beam spectrograph on the 5-m Hale Telescope at Palomar Observatory. Our spectrum complements that presented in \citet{tran00}; their spectrum was taken in 112 minutes using LRIS on Keck, and so is deeper, but our spectrum extends $\sim700$\AA\ further redward and was taken closer in time to the X-ray data. The night was photometric, albeit with $2\arcsec$ seeing. We observed \iras\ for 500~s, split into two equal exposures.  We used the 5500~\AA\, dichroic, the $2\arcsec$ wide longslit, the 600 $\ell\, {\rm mm}^{-1}$ grating on the blue arm of the spectrograph (blazed at 4000~\AA; resolving power $R \equiv \lambda / \Delta \lambda \sim 900$), and the 316 $\ell\, {\rm mm}^{-1}$ grating on the red arm of the spectrograph (blazed at 7500~\AA; $R \sim 1200$). The data were processed using standard procedures within the Image Reduction and Analysis Facility (IRAF) environment. Flux-calibration was calculated using observations of standard stars Feige~66 and Feige~110 from \citet{massey90}, obtained on the same night.

\subsection{X-ray}\label{obsxray}
\iras \ was observed with \nustar \ \citep{harrison13} on 2012 December~26 with a total exposure of 15.2~ks (OBSID 60001067) as part of the \nustar\ program to observe type 2 QSOs \citet{lansbury+2014,lansbury+2015}. The observation was coordinated with the \swift \ observatory, which observed the same target on 2012 December~25 (OBSID 00080413001). The total \swiftxrt \ exposure was 6.0~ks. The \nustar \ and \swift \ observations are sufficiently close in time that they provide a quasi-simultaneous snapshot of \iras \ across the broad 0.5--70~keV band. This observing strategy is typical for the \nustar \ snapshot survey of AGN in the nearby Universe (Balokovi\'{c} et al., in preparation). The \swift \ and \nustar \ data are presented here for the first time. All observations used in this paper are listed in Table~\ref{xrayobstable}.

The \nustar \ data were reduced in the manner described in \citet{perri+2014}. We used HEASOFT~v\,6.16, NuSTARDAS~v\,1.4.1, and CALDB version 20150316, with a 50\arcsec\ extraction radius. Following the event filtering, we extracted the source spectrum from a circular aperture centered on the peak of the point source. The background extraction region covered the free area of the same detector, excluding a region of $\simeq$80\arcsec\ radius around the source. The target is detected in the 10--50~keV band with signal-to-noise ratio of $\simeq$10 in FPMA, and $\simeq$8 in FPMB. The 10--50~keV (3--79~keV) background-subtracted count rates are 0.010~s$^{-1}$ (0.034~s$^{-1}$) and 0.008~s$^{-1}$ (0.032~s$^{-1}$). The spectrum and corresponding response files were generated using the {\tt nuproducts} script. Spectra for each \nustar \ module are binned to a minimum of 20 counts per bin, and fitted simultaneously as described in \S\ref{analysis_xray}. We allowed the cross-normalization factor to vary in all fits (with instrumental normalization of FPMA fixed at unity), and found it to be consistent with unity to within 5\,\% in all cases.

We used resources provided by the ASDC\footnote{\url{http://www.asdc.asi.it/mmia/}} for \swiftxrt \ data reduction. The spectrum was extracted from a region with a radius of 20\arcsec\ centered on the brightest peak of emission, and the background was sampled from an annulus extending between 40\arcsec\ and 80\arcsec\ around the source. For spectral fitting we used the source spectrum binned to a minimum of 20~counts per bin before background subtraction. The \swift \ data photon statistics are well matched to those of the \nustar \ data. 

We complement these data with archival X-ray data taken with \suzaku\ on 2011 November~18 (OBSID 706038010; 81~ks), and with \chandra\ on 2009 January~6 (OBSID 10445; 69~ks). We largely followed the processing steps of \citet{chiang+2013} for these datasets in order to facilitate a direct comparison of the results, so we refer the reader to their data section for details. The \suzaku\ data were reduced using standard procedures\footnote{\url{http://www.astro.isas.jaxa.jp/suzaku/process/}}. No detection was achieved with the HXD/PIN, so we only made use of the soft X-ray data. The spectra were extracted from circular regions 100\arcsec\ in radius, which includes most of the diffuse emission. Background spectra were extracted from emission-free areas of each XIS detector. The spectra from the two front-illuminated chips (XIS0 and XIS3) were coadded. We binned the spectra to a minimum signal-to-noise ratio of~3 and ignored any data outside of the 0.5--8.5~keV range. 

The \chandra\ data were processed using CIAO version 4.6. We extracted the nuclear spectrum from a circular region 1\arcsec\ in radius centered on the peak of the emission. Background was sampled from a ring with an inner radius of 2\arcsec\ and an outer radius of 4\arcsec; in this way most ($\gtrsim90\,\%$) of the diffuse emission contribution to the unresolved central source is removed. Unlike all other instruments used in this work, which sample both the AGN and diffuse emission on $\sim$10\arcsec\ scales, \chandra\ allows us to isolate the AGN-dominated core emission. In order to assess the contribution of diffuse emission in \nustar\ and \suzaku\ apertures, we also extracted \chandra\ spectra from circular regions with 50\arcsec\ and 100\arcsec\ radii. These extractions are used only in the comparison between instruments presented in \S\,\ref{sec:xray_diffuse}.

\begin{deluxetable}{lcc}
\tabletypesize{\scriptsize}
\tablecolumns{3}
\tablewidth{0pc}
\tablecaption{Emission line properties of \iras\ measured from the Palomar Double Spectrograph.}
\tablehead{\colhead{Line}&\colhead{Flux}&\colhead{Rest EW}\\
\colhead{}&\colhead{$(10^{-14}$erg cm$^{-2}$ s$^{-1})$}&\colhead{(\AA)}}
\startdata
$[$\ion{S}{2}$]$                  6734 &   2.24 $\pm$ 0.15 &  58  $\pm$ 20 \\
$[$\ion{S}{2}$]$                  6716 &   2.63 $\pm$ 0.15 &  62  $\pm$ 15 \\
$[$\ion{N}{2}$]$                  6583 &   6.44 $\pm$ 0.62 & 141  $\pm$ 15 \\
H$\alpha$                         6563 &   8.24 $\pm$ 0.44 & 183  $\pm$ 14 \\
$[$\ion{N}{2}$]$                  6543 &   3.47 $\pm$ 0.80 &  78  $\pm$ 11 \\
$[$\ion{Fe}{10}$]$                6374 &   0.60 $\pm$ 0.20 &  17  $\pm$ 10 \\
$[$\ion{O}{1}$]$+$[$\ion{S}{3}$]$ 6300 &   1.42 $\pm$ 0.30 &  40  $\pm$ 17 \\
$[$\ion{Fe}{7}$]$                 6087 &   0.77 $\pm$ 0.27 &  23  $\pm$ 13 \\
\ion{He}{1}           			  5876 &   0.48 $\pm$ 0.15 &  15  $\pm$ 8  
\enddata  
\tablecomments{A higher resolution, deeper optical spectrum is available in \citet{tran00}. We here present those lines that are uniquely present in our spectrum due to our longer wavelength coverage, plus two lines in the wavelength range in which our data overlap with \citet{tran00}. See also \citealt{craw96,soi96}.}
\label{optfluxes}
\end{deluxetable}

\begin{deluxetable}{lccc}
\tabletypesize{\scriptsize}
\tablecolumns{4}
\tablewidth{0pc}
\tablecaption{X-ray observations of \iras\ used in this paper}
\tablehead{\colhead{Observatory}&\colhead{Observation}&\colhead{Exposure}&\colhead{Source Count}\\
\colhead{and Instrument}&\colhead{Start Date}&\colhead{(ks)}&\colhead{Rate ($10^{-2}$\,s$^{-1}$)}}
\startdata
\nustar/FPMA   & 2012-Dec-26 & 15.2 & $3.4\pm0.2$ \\
\swiftxrt      & 2012-Dec-25 & 5.9  & $5.0\pm0.3$ \\
\suzaku/XIS1   & 2011-Nov-18 & 81.3 & $12.6\pm0.1$\\
\chandra/ACIS  & 2009-Jan-06 & 69.3 & $1.09\pm0.04$\\
\enddata  
\tablecomments{Count rates are background-subtracted rates for one of the instruments of a given observatory, within the source extraction region and bandpass used for fitting (see \S\,\ref{obsxray} and \S\,\ref{analysis_xray} for details).}
\label{xrayobstable}
\end{deluxetable}

\section{Infrared \& Optical Analysis}\label{iroptanalysis}
We assume that the IR emission arises from a single episode of star formation and/or AGN activity. We then fit the IR data simultaneously with two grids of pre-computed radiative transfer models; one for dusty AGN \citep{efst95,efst13} and one for starbursts \citep{efst00}. A model set for old stellar populations is nor included, for the reasons given in \S\ref{iroptobs}. These models have been used previously in, e.g., \citet{ver02,farrah02b,farrah03,farrah12,efst13}. The AGN models assume the dust geometry is a smooth tapered disk whose height, $h$, increases linearly with distance, $r$, from the AGN until it reaches a constant value. The dust distribution includes multiple species of varying sizes, and assumes the density distribution scales as $r^{-1}$. The AGN model parameters are: inner half-opening angle of the torus measured from pole-on ($15\degr - 60\degr$), viewing angle measured from pole-on ($1\degr - 90\degr$), ratio of inner to outer disc radius ($r_1/r_2 = 0.00625-0.05$), ratio of height to outer radius ($h/r_2 = 0.0625-0.5$), and equatorial optical depth at 1000\AA\ (250 to 1250, defined in equations 1 \& 2 of \citealt{efst95}, see also \citealt{efst90}). The starburst models combine the stellar population synthesis code of \citet{bru03} with a prescription for radiative transfer through dust that includes the effects of small dust grains and polycyclic aromatic hydrocarbons (PAHs) updated with the dust model of \citet{efst09}. The starburst model parameters are age (0--70\,Myr), initial optical depths of the molecular clouds ($\tau_{V}$ = 50, 75, and 100), and $e$-folding timescale for the starburst ($10-40$\,Myr). In total there are 1680 starburst models and 4212 AGN models. 

\begin{figure} 
\begin{center}
\includegraphics[width=0.75\columnwidth,angle=90]{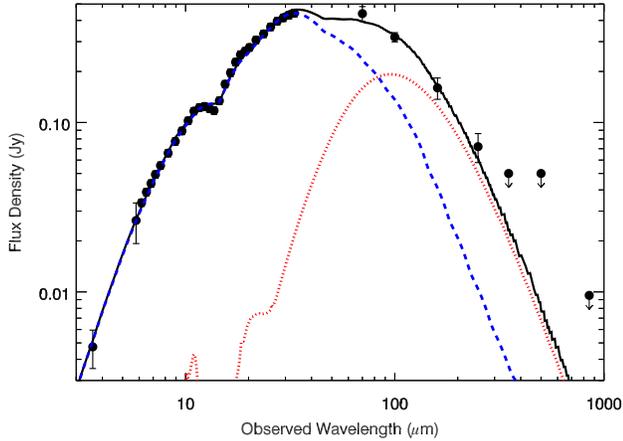} 
\caption{The best-fit ($\chi^2_{red}=0.7$) IR spectral energy distribution for \iras. The black line is the combined model, the blue line is the AGN, and the red line is the starburst. The data include IRAC photometry, IRS spectroscopy (see also \citealt{zakam08,sar08,shache12,ruiz13}), PACS, SPIRE, and SCUBA photometry.}
\label{fig:irsed}
\end{center}
\end{figure}

The best-fit IR SED is shown in Figure \ref{fig:irsed}. The total IR (rest-frame 1-1000$\mu$m) luminosity is $6.76\pm0.20\times10^{46}$\,erg\,s$^{-1}$, with a contribution from the AGN of $5.94^{+0.26}_{-0.27}\times10^{46}$\,erg\,s$^{-1}$. The starburst is required in the fit at $3.7\sigma$ confidence, with a luminosity of $5.54\pm1.48\times10^{45}$\,erg\,s$^{-1}$, corresponding to a star formation rate of $(110^{+35}_{-28})$\,M$_{\odot}$ yr$^{-1}$. The uncertainties on these parameters are the 68\% confidence intervals, evaluated using the method in \citet{farrah12}. 

The combination of a mid-IR spectrum with far-IR photometry allow constraints to be set on other model parameters. Since the IR data are however still relatively limited, we have deduced these constraints by considering all the individual solutions in weighted probability distribution functions, and so do not consider how these constraints may depend on each other. We have also not explored how these constraints depend on the choice of model set. In particular, we have not explored how these constraints may change if a clumpy, rather than smooth, dust distribution is assumed. With these caveats in mind, we present the following results. The starburst age is constrained (at 3$\sigma$) to be $<50$\,Myr. The line of sight viewing angle, $\theta^{\rm ir}_{\rm V}$ to the IR-emitting torus is $(35^{+8}_{-5})$\degr. The half opening angle of the torus, $\theta^{\rm ir}_{\rm L}=(36^{+9}_{-6})$\degr, is indistinguishable from  $\theta^{\rm ir}_{\rm V}$. The inner to outer radius ratio of the torus is $0.016^{+0.006}_{-0.004}$, while the ratio of the torus height to the outer radius is $0.16^{+0.06}_{-0.04}$. We compare these values to those previously reported in the literature \citep{hines93,hines99,tran00,burts13} in Sections \ref{discir} and \ref{discalldat}.

The torus geometry assumed in the \citealt{efst95} models means the mid-IR emission is anisotropic, with viewing angles closer to edge-on tending to suppress the mid-IR emission \citep{efst06,efst14}. The derived combination of torus geometry and viewing angle of \iras\ thus imply a (multiplicative) anisotropy correction to the AGN luminosity of $0.83^{+0.08}_{-0.07}$. The derived {\itshape intrinsic} AGN infrared luminosity is thus $\sim4.9\times10^{46}$\,erg\,s$^{-1}$ and a total IR luminosity (assuming the starburst emission is isotropic) of $\sim5.5\times10^{46}$\,erg\,s$^{-1}$. 

\begin{figure} 
\begin{center}
\includegraphics[width=0.75\columnwidth,angle=-90]{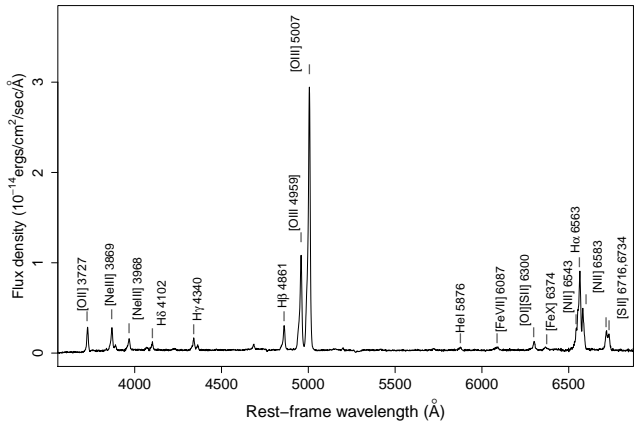}
\caption{Optical spectrum of \iras \ taken with the Double Spectrograph at Palomar Observatory. Together with the canonical emission lines there are weak but clear detections of two coronal iron lines. Line fluxes are given in Table \ref{optfluxes}.}
\label{fig:opspec}
\end{center}
\end{figure}

The optical spectrum is shown in Figure \ref{fig:opspec}. From it, we derive $z=0.4416\pm0.0001$. The optical spectrum shows multiple narrow emission lines (Table~\ref{optfluxes}, see also \citet{tran00}). Analysis of the spectrum was conducted within IRAF. Lines were identified using line lists assembled from previous observations of starbursts and AGN \citep{far05,shir12}. Line fluxes and equivalent widths were measured by marking two continuum points, one on each side of the line, and fitting a linear continuum. The errors were found by estimating the variance in the continuum and remeasuring the EW using the variance as the continuum level.

Our spectrum is consistent within the errors with that of \citet{tran00}; given that their spectrum is of higher quality than ours we only comment on the lines in the additional wavelength coverage of our spectrum. We detect the canonical emission lines, including H$\alpha$, [\ion{N}{2}] and [\ion{S}{2}] lines. From our spectrum, \iras\ is unambiguously classified as a Seyfert using standard emission-line diagnostics \citep{bpt81}, and lies well away from the regions proposed as harboring composite AGN/starburst systems \citep[e.g.][]{kewley01,stasinska06}. In addition we detect two high excitation `coronal' iron lines: [\ion{Fe}{7}]$\lambda$6087 and [\ion{Fe}{10}]$\lambda$6375, at $2.8\sigma$ and $2.9\sigma$ significance, respectively. These iron lines have been seen in ULIRGs \citep{far05} but are more commonly observed in supernova remnants and in the Solar corona (hence their name). They are rare in extragalactic objects (but see \citealt{ost81,rey,gel09,rose15} for examples). They are discussed further in Section \ref{discalldat}.

\section{X-ray Analysis}\label{analysis_xray}

\subsection{Overview}\label{xraymodover}
We use \xspec\,\citep{arnaud-1996} version 12.8.2 for spectral modeling of the \nustar, \swift, \suzaku, and \chandra\ data. To model the soft X-ray data we follow results from previous studies and use two \mekal\ plasma components \citep{mew85,lie95} with temperatures determined directly from the data. In the fits to the (joint) data we keep the temperatures the same for all instruments but allow normalizations to vary independently, because different instruments' point spread functions sample the diffuse emission differently, and spatial variation in temperature has previously been found \citep{osullivan+2012}. Other \mekal\ parameters are kept fixed at the values determined in previous work: $n_{\rm H}=5$~cm$^{-3}$, and $Z=0.4\,Z_{\odot}$. Since much of the diffuse emission is subtracted from the small-scale \chandra\ spectrum, we set the normalization of one of the two \mekal\ components to zero for these data, and model the residual plasma contribution with the remaining \mekal\ component. Due to the limited quality of the \swiftxrt\ spectrum, we also use only a single \mekal\ component to model it. All models include Galactic absorption ($N_{\rm H,G}=1.4\times10^{20}$~cm$^{-2}$; \citealt{kalberla+2005}), and the same redshift, $z=0.442$, for all components.

To model the hard X-ray data from \nustar, \swift, \suzaku, and \chandra\ we use two model sets:

\begin{enumerate}

  \item T$+$R -- a phenomenological model consisting of two independent components, one transmitted (T) and one reflected (R). The T component is an absorbed power law modeled by \xspec\ model {\tt wabs}$\times${\tt cabs}$\times${\tt cutoffpl}, which accounts for Compton scattering and has a fixed $e$-folding scale of 200~keV. The R component is modelled using \pexrav\ \citep{magdziarz+zdziarski-1995}, and a narrow Gaussian emission line at 6.4~keV representing fluorescent iron\,K$\alpha$ emission.

  \item Torus models -- observationally motivated geometric models, in which the T \& R components are self-consistently calculated and coupled. We consider two torus models, \mytorus\ \citep{murphy+yaqoob-2009} and \bntorus \citep{brightman+nandra-2011}.

\end{enumerate}

The T$+$R model has been used by many previous authors, so we employ it to allow for straightforward comparisons. It was predominantly used in one of two extremes, transmission-dominated (TD) and reflection-dominated (RD), where one of the components was assumed to be negligible. However, the key to insights into the properties of the X-ray obscurer is the ability to model both components \citep{yaqoob-2012}, unless the obscuration is so high ($N_{\rm H}\gtrsim10^{25}$~cm$^{-2}$) that only the R component is observable \citep[e.g.][]{arev14,balokovic+2014,gandhi+2014,annuar+2015,bauer15}. Here we start with two components and let the data determine if either component is negligible. This model does not have a physical geometry, but has nevertheless been used in the literature to account for spectral features attributed to the AGN torus. In particular, \pexrav\ assumes a slab geometry rather than a torus, so the viewing angle changes the spectrum at the level of only a few percent over most of the 0--90\degr\ range. The viewing angle is kept fixed ($\cos \theta_{\rm V}=0.45$) because it cannot be interpreted in the context of the torus, so it should not be compared to other viewing angles discussed in this paper. This model also includes an unresolved Gaussian line fixed at $E=6.4$~keV ($\sigma=10^{-3}$~keV), accounting for fluorescent emission of iron arising from the same material producing the $R$ component. We keep the elemental abundances in \pexrav\ fixed at Solar values, and the normalization of the 6.4~keV line independent of the \pexrav\ normalization.

In contrast, \mytorus\ and \bntorus\ are models for the obscurer with an observationally motivated geometry, that of a smooth toroidal obscurer. The geometry assumed in the \mytorus\ model is a torus with a fixed inner half-opening angle of $\theta^{\rm my}_{\rm L} = 60$\degr. The column density in the line of sight {($N_{\rm H}$) is a function of viewing angle and varies from maximum for a viewing directly through the equator (where $N_{\rm H}=N_{\rm H,eq}$)} to zero when the line of sight just grazes the torus edge. In the \bntorus\ model the torus is approximated as a sphere with symmetric conical cutouts and the inner half-opening angle $\theta^{\rm bn}_{\rm L}$ can be varied as a fitting parameter. The column density has a single value along any line of sight that intersects the torus; that is, as long as $\theta^{\rm bn}_{\rm V}>\theta^{\rm bn}_{\rm L}$ then the line-of-sight column density $N_{\rm H}$ is equal to the equatorial column density $N_{\rm H,eq}$. Since the normalizations of different spectral components are internally linked due to the obscurer geometry, two degrees of freedom ($\nu$) are removed from the fits with respect to the T$+$R model. In both torus models the Fe\,K$\alpha$ line strengths are self-consistently calculated.

We describe the  T$+$R model fits in \S\ref{tprsection} and the torus modeling in \S\ref{tormod}. The parameters of the X-ray models for the AGN are summarized in Table~\ref{xraymodtable}. We consider models for the diffuse emission separately in \S\ref{sec:xray_diffuse}.

\begin{figure} 
\begin{center}
\includegraphics[width=0.95\columnwidth]{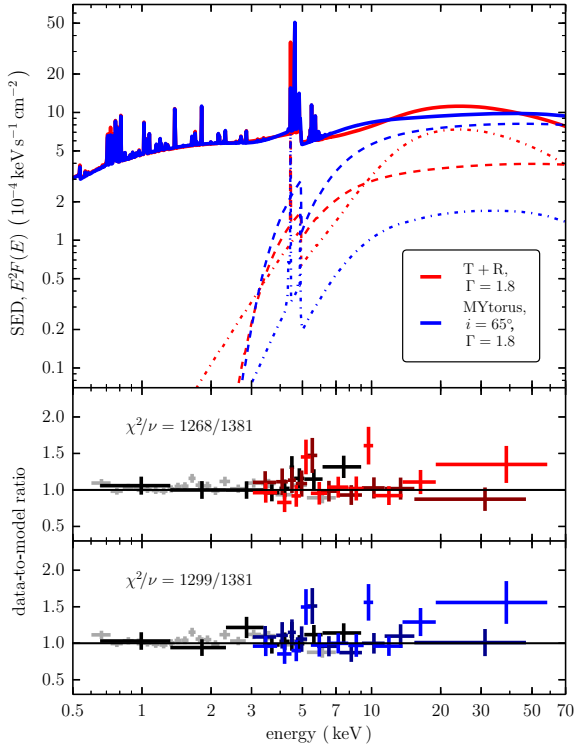}
\caption{Comparison of best-fit models for the \nustar, \suzaku\ and \chandra\ data. A phenomenological T$+$R model is shown by the red lines, while blue lines show a \mytorus\ model. Solid lines are for the total spectrum (AGN and diffuse emission), dashed lines are for the transmission components, and dot-dashed lines are for the reflection components. Plasma components making up the diffuse emission are not plotted in order to avoid confusion; any flux not contributed by the AGN components is due to plasma emission. The lower panels show the data-to-model ratio for each of the two models. Colored lines are for the \nustar\ data (darker color for FPMA, brighter color for FPMB), grey is for \suzaku, and black is for \chandra.}
\label{fig:xray_fig1}
\end{center}
\end{figure}

\subsection{The T$+$R Model}\label{tprsection}
We first model only the simultaneous \swift\ and \nustar\ spectra, which are well matched in signal-to-noise ratio across the broad 0.5--50\,keV bandpass. Fitting the T$+$R model, we find that the photon index, $\Gamma$, cannot be constrained. Any photon index in the range $1.4 < \Gamma < 2.6$ fits the data equally well as the canonical $\Gamma=1.8$ (e.g., \citealt{dadina-2008,rivers+2013,malizia+2014}). Fixing $\Gamma$ at 1.8, the best fit ($\chi^2/\nu=76.7/63$) is for a model with the intrinsic power-law continuum absorbed by $N_{\rm H}\sim4\times10^{24}$~cm$^{-2}$, with contributions from both T and R components. However, valid solutions exist with no absorbed component present. In the best-fit solution, the T component dominates at energies above 20~keV. Assuming a harder photon index ($\Gamma\approx1.6$) leads to TD solutions, while a softer assumed index ($\Gamma\approx2.1$) gives RD solutions. In either case, $\chi^2$ increases by less than 1 with respect to the best fit. The diverse range of models consistent with this dataset constrains the intrinsic 2--10~keV luminosity to lie between $4\times10^{45}$~erg\,s$^{-1}$ and $1\times10^{47}$~erg\,s$^{-1}$.

To provide more stringent constraints on the models, we model the \nustar\ data (taken December 2012) together with archival \suzaku\ and \chandra\ data (taken November 2011 and January 2009, respectively). The \swift, \suzaku\ and \chandra\ data are consistent with each other, but because of poorer photon statistics we exclude the \swift\ data from modeling. With the additional \suzaku\ and \chandra\ data, the constraints on the photon index and the absorption column improve significantly. Note that in this case, the very high signal-to-noise ratio of the soft X-ray data constrains models better than the \nustar\ data in the overlapping energy range.

\begin{figure} 
\begin{center}
\includegraphics[width=0.95\columnwidth]{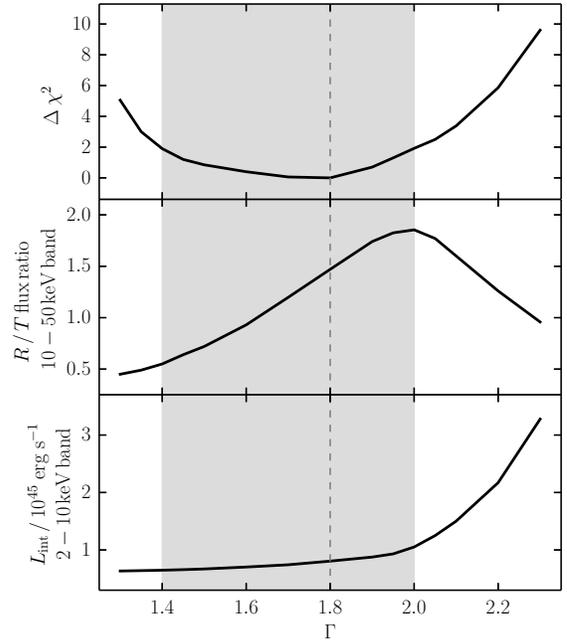}
\caption{Phenomenological description of the AGN spectrum of \iras\ as a function of the intrinsic photon index. For $\Gamma\approx1.4$ the 10--50~keV band is dominated by the T~component, while for $\Gamma\approx2.0$ the R~component dominates, as shown in the middle panel. The joint \nustar, \suzaku\ and \chandra\ data constrain the photon index to $\Gamma = 1.8_{-0.4}^{+0.2}$, marked with the grey shaded area and the vertical dashed line. Within this confidence interval, the intrinsic 2--10~keV luminosity is well constrained, as shown in the bottom panel. For comparison, the fits based only on the simultaneous \nustar\ and \swiftxrt\ data transition from being dominated by the T component to being dominated by the R component within a narrower range of $1.7<\Gamma<2.0$. In that case, the intrinsic luminosity spans more than an order of magnitude within the range of $\Gamma$ plotted here, while the change in $\chi^2$ barely exceeds unity over that parameter range.}
\label{fig:xray_fig2}
\end{center}
\end{figure}

We find the best fit ($\chi^2/\nu=1268/1381$) for $\Gamma=1.8_{-0.4}^{+0.2}$ and $N_{\rm H}=\left(5_{-2}^{+3}\right)\times10^{23}$~cm$^{-2}$. This model is shown in Figure~\ref{fig:xray_fig1}. The soft X-ray data alone drive the fit toward hard photon indices ($\Gamma<1.5$) and a TD model \citep{chiang+2013}. The addition of \nustar\ data constrains $\Gamma$ to a more typical value and results in a solution where T and R components contribute to the hard X-ray flux nearly equally. Figure~\ref{fig:xray_fig2} illustrates how the $\chi^2$, the relative contributions of T and R components, and the implied intrinsic luminosity vary within the 90\% confidence interval for the photon index ($1.4-2.0$). The intrinsic 2--10~keV luminosity of the best-fit model is $8\pm3\times10^{44}$~erg\,s$^{-1}$.

Additional constraints can be drawn from the equivalent width (EW) of the neutral Fe\,K$\alpha$ line\footnote{We evaluate the equivalent width of the Fe\,K$\alpha$ line by taking the ratio of line flux to flux density of the AGN continuum components only, i.e., excluding the plasma components that otherwise dominate up to the Fe\,K$\alpha$ line energy at 6.4~keV, except for the small-scale \chandra\ spectrum. We use a band spanning rest-frame 5.7--6.7 keV.}. The low equivalent width of this line ($EW\simeq0.3$~keV) in the \nustar, \swift, and archival data, argues against an RD scenario, since RD spectra usually have Fe\,K$\alpha$ EWs of $\sim\,$1~keV. On the other hand, a weak iron line could also arise if the iron abundance is $\sim$30\% Solar, which is plausible given that sub-Solar abundance has been inferred for the diffuse plasma (e.g., \citealt{osullivan+2012}). Constraints based on the Fe\,K$\alpha$ line are discussed in \citet{chiang+2013}; due to the inferior spectral resolution of \nustar\ compared to \chandra\ around 6.4~keV ($\simeq 0.4$~keV compared to $\simeq 0.13$~keV), the new data do not alter their conclusions.

\subsection{The Torus Models}\label{tormod}
We start by applying the \mytorus\ model to the simultaneous \swift\ and \nustar\ data. Due to limited photon statistics we can only draw tentative conclusions. If we fix $\Gamma$ to 1.8 and assume that the torus is viewed edge-on ($\theta^{\rm my}_{\rm V}=90^{\circ}$) then a good fit ($\chi^2/\nu=78.8/65$) is found for $N_{\rm H}=\left(2_{-1}^{+4}\right)\times10^{24}$~cm$^{-2}$. In this case the equatorial column density of the torus, $N_{\rm H,eq}$, equals the column density observed along the line of sight to the nuclear X-ray source, $N_{\rm H}$. If we let viewing angle vary then this implies lower $N_{\rm H}$, but for $\theta^{\rm my}_{\rm V}<75\degr$ the fits only produce a lower limit on $N_{\rm H,eq}$ of about $3\times10^{24}$~cm$^{-2}$. There is perhaps a slight preference for viewing angles closer to $\theta^{\rm my}_{\rm V}=60\degr$ (i.e., the edge of the torus in the \mytorus\ model), but the corresponding change in $\chi^2$, relative to edge-on inclination, is less than 2. The solutions are generally RD, with EWs of Fe\,K$\alpha$ of $\leq1.2$~keV. The implied intrinsic luminosity in the 2-10 keV band is $9\pm2\times10^{45}$~erg\,s$^{-1}$. 

Applying the \mytorus\ model to the joint \nustar, \suzaku, and \chandra\ dataset, we find a preference away from edge-on inclination. Due to the geometry assumed in the model, viewing angles within $\sim5\degr$ of $60\degr$ (where the line of sight skims the torus) require caution, as $N_{\rm H}$ changes steeply with viewing angle -- this can lead to unreasonably tight constraints on some model parameters. We therefore fix $\theta^{\rm my}_{\rm V}$ to $65\degr$. The best fit is an effectively TD model, with the R component contributing $\lesssim$20\% to the 10--50~keV band and an Fe\,K$\alpha$ EW of $0.3\pm0.1$~keV. This is consistent with the low Fe\,K$\alpha$ EW found with the same data using the T$+$R model. The equatorial column density of the torus, $N_{\rm H,eq}$, is $\left(9\pm2\right)\times10^{23}$~cm$^{-2}$ for $\Gamma$ in the range $1.6-1.8$. The best-fit photon index is 1.6, but this is at the lower end of the parameter domain for the \mytorus\ model, so a true lower limit to the confidence interval cannot be determined. Assuming a statistically acceptable value of $\Gamma=1.8$ leads to $N_{\rm H,eq}=\left(1.1_{-0.1}^{+0.2}\right)\times10^{24}$~cm$^{-2}$. The intrinsic 2--10~keV luminosity from this model is $(1.1\pm0.1)\times10^{45}$~erg\,s$^{-1}$. This fit is shown in Figure~\ref{fig:xray_fig1}.

Applying the \bntorus\ model to the \swift\ and \nustar\ data, we find an equally good fit as the \mytorus\ model. Assuming $\Gamma=1.8$, the best fit ($\chi^2/\nu=73.9/64$) is found for $N_{\rm H}>2\times10^{24}$~cm$^{-2}$. Again, this is an RD solution, with the $N_{\rm H}$ constrained from the upper side only by the parameter domain limit ($<10^{26}$~cm$^{-2}$). We find that $\theta^{\rm bn}_{\rm V}$ and $\theta^{\rm bn}_{\rm L}$ cannot be constrained by the data simultaneously; however, fixing $\theta^{\rm bn}_{\rm L}$ always leads to $\theta^{\rm bn}_{\rm V}$ lying between $\gtrsim\left(\theta^{\rm bn}_{\rm L}\right)$\degr and edge-on. The implied 2--10~keV intrinsic luminosity lies between $8\times10^{45}$~erg\,s$^{-1}$ and $2\times10^{46}$~erg\,s$^{-1}$. 

The \bntorus\ model applied to the joint \nustar, \suzaku, and \chandra\ data does not provide simultaneous constraints on $\theta^{\rm bn}_{\rm V}$ and $\theta^{\rm bn}_{\rm L}$ either. They are constrained in the sense that their difference is $\gtrsim5^{\circ}$ for any one assumed angle within their respective parameter ranges; 18--87$^{\circ}$ for $\theta^{\rm bn}_{\rm V}$ and 26--84$^{\circ}$ for $\theta^{\rm bn}_{\rm L}$, which is consistent with the Sy2 classification (i.e. that the {\itshape optical} BELR is not seen in direct light). We find best fits ($\chi^2/\nu=1276/1379$) consistent with $\Gamma=1.6\pm0.2$ and $N_{\rm H}=\left(4\pm1\right)\times10^{23}$~cm$^{-2}$ for a broad range of viewing angles. Intrinsic 2--10~keV luminosities for these solutions are in the range $(1.2-1.8)\times10^{45}$~erg\,s$^{-1}$. Although the T and R components, as well as iron lines, cannot be separated in this model, equivalent phenomenological solutions reveal that the T component dominates the $>10$~keV flux. The best-fit solution is therefore qualitatively similar to that obtained from the \mytorus\ model.

\begin{deluxetable}{lcc}
\tabletypesize{\scriptsize}
\tablecolumns{3}
\tablewidth{0pc}
\tablecaption{Summary of modeling of the X-ray spectrum}
\tablehead{\colhead{Model}&\multicolumn{2}{c}{Data:~\nustar\ with} \\
\colhead{Parameter}&\colhead{\swiftxrt}&\colhead{\suzaku\ and \chandra}}
\startdata
\cutinhead{T$+$R model} 
$\chi^2/$d.o.f. & 76.7/63\,\tablenotemark{a} & 1268/1381 \\
$\Gamma$ & $\left[1.6,\,2.1\right]$ & $1.8_{-0.4}^{+0.2}$ \\
$L_{2-10\,{\rm keV}}$ & $\left[4,\,100\right]$ & $0.8\pm0.3$\\
$N_{\rm H}$ & $>4$ & $0.5_{-0.2}^{+0.3}$ \\
\cutinhead{\mytorus\ model} 
$\chi^2/$d.o.f. & 78.8/65 & 1299/1381 \\
$\Gamma$ & 1.8 (f) & $<1.8$\,\tablenotemark{b} \\
$L_{2-10\,{\rm keV}}$ & $9\pm2$ & $1.1\pm0.1$\\
$N_{\rm H,eq}$ & $2_{-1}^{+4}$ & $0.9\pm0.2$\,\tablenotemark{c} \\
$N_{\rm H}$ & $=N_{\rm H,eq}$ & $0.5\pm0.1$ \\
$\theta_{\rm V}^{\rm my}$ & 90 (f) & 65 (f) \\
\cutinhead{\bntorus\ model\,\tablenotemark{d}} 
$\chi^2/$d.o.f. & 73.9/64 & 1276/1379 \\
$\Gamma$ & 1.8 (f) & $1.6\pm0.2$ \\
$L_{2-10\,{\rm keV}}$ & $\left[8,20\right]$ & $\left[1.2,1.8\right]$ \\
$N_{\rm H,eq}=N_{\rm H}$ & $>2$ & $0.4\pm0.1$ \\
$\theta_{\rm V}^{\rm bn}$ & $\left[\theta_{\rm L}^{\rm bn},90\right]$ & $\left[\theta_{\rm L}^{\rm bn}+5,\theta_{\rm L}^{\rm bn}+15\right]$\\
\enddata  
\tablecomments{Spectral parameters are: intrinsic photon index ($\Gamma$), intrinsic 2--10\,keV luminosity ($L_{2-10\,{\rm keV}}$, in units of $10^{45}$~erg\,s$^{-1}$), line-of-sight column density ($N_{\rm H}$, in $10^{24}$\,cm$^{-2}$), equatorial column density ($N_{\rm H,eq}$, in $10^{24}$\,cm$^{-2}$), viewing angle, ($\theta_{\rm V}$, in deg.), and torus half-opening angle ($\theta_{\rm L}$, in deg.). Numbers in square brackets denote ranges and fixed parameters are followed by (f).}
\tablenotetext{a}{Evaluated for $\Gamma=1.8$ and $N_{\rm H}=4$ (in the same units); $\Delta\chi^2<1$ for the parameters' ranges shown here.}
\tablenotetext{b}{Best fit is $\Gamma=1.6$, which is the edge of the parameter domain for $\Gamma$.}
\tablenotetext{c}{For $\Gamma=1.8$, $N_{\rm H,eq}=1.1_{-0.1}^{+0.2}$ (in the same units).}
\tablenotetext{d}{Since angles $\theta_{\rm V}^{\rm bn}$ and $\theta_{\rm L}^{\rm bn}$ cannot be constrained independently, we express constraints on $\theta_{\rm V}^{\rm bn}$ in terms of $\theta_{\rm L}^{\rm bn}$.}
\label{xraymodtable}
\end{deluxetable}

\subsection{Diffuse Emission Models\\and Multi-epoch Flux Comparison}\label{sec:xray_diffuse}
In the modeling presented in \S\ref{tprsection} and \S\ref{tormod}, the diffuse emission was included but the focus was on the AGN emission. It is however worth briefly discussing the diffuse emission models, for two reasons. First, while the literature is consistent in modeling the extended emission, details and best-fit parameters differ between studies. Second, with relatively high temperatures, the diffuse plasma emission significantly contributes to the emission into the \nustar\ band, up to $\simeq10$~keV. 

As the extended emission exhibits significant spatial variations in plasma temperature \citep{osullivan+2012}, a cross-instrument comparison based on a simple one- or two-component \mekal\ model is only approximate. However, we did not find it necessary to add complexity to the model based on fitting statistics or structured residuals. We find best-fit plasma temperatures in the range of 1--4~keV and 5--8~keV, based mostly on \nustar\ and \suzaku\ data. For any single model fit (recalling that we only use a single \mekal\ model for both the \chandra\ and \swiftxrt\ data, see \S\ref{xraymodover}), the typical 90\% uncertainty on the temperature is 0.4--1.5~keV when \nustar\ is combined with the archival data, and approximately 2~keV when combined with \swiftxrt. These results are similar to all previous studies.

The most direct comparison can be made between \nustar\ and \chandra\ spectra extracted from the same 50\arcsec\ circular region. In this case we find that the total flux in the 3--8~keV band is $(1.0\pm0.1)\times10^{-12}$~erg\,s$^{-1}$\,cm$^{-2}$ in both instruments. Assuming that the \swiftxrt\ extraction contains most of the diffuse emission, its 3--8~keV flux of $9.3\times10^{-13}$~erg\,s$^{-1}$\,cm$^{-2}$ is also consistent with \nustar\ within the typical spread found in other simultaneous observations. The cross-normalization between the two modules of \nustar, as well as those of \suzaku, is within 5\% of unity in all models. We thus find that no significant spectral variability occured between \chandra, \suzaku, \nustar\ and \swift\ observations, and that all cross-normalizations discussed here are well within their respective expectations \citep{madsen+2015}.

The extended soft X-ray emission spans several tens of kpc \citep{osullivan+2012},
and therefore should not vary on a timescale spanning the observations used here. We confirm this based on spectra extracted from large circular regions (100\arcsec\ for \suzaku\ and \chandra). The small-scale \chandra\ spectrum (within 1\arcsec) is dominated by AGN emission above 3~keV according to nearly all models, with a flux in the 3--8~keV band of $(4.0\pm0.2)\times10^{-13}$~erg\,s$^{-1}$\,cm$^{-2}$. In the \nustar\ spectra, the diffuse emission contributes approximately 10\% of the flux even at 10~keV.

The 3--8~keV flux from the best-fit AGN components in different models ranges from $3.6\times10^{-13}$~erg\,s$^{-1}$\,cm$^{-2}$ to $3.9\times10^{-13}$~erg\,s$^{-1}$\,cm$^{-2}$, which is consistent with the nuclear \chandra\ flux. With our 2--10~keV flux of $1.2-3.6\,\times 10^{-13}$~erg\,s$^{-1}$\,cm$^{-2}$ (AGN components alone, based on the \nustar\ data), we find excellent agreement with the flux estimated by \citet{chiang+2013} assuming two different AGN models based on \chandra\ and \suzaku\ data, ranging over $1.8-3.3\,\times 10^{-13}$~erg\,s$^{-1}$\,cm$^{-2}$. The \xmm-based estimate of \citet{piconcelli+2007}, $4.7-5.5\,\times 10^{-13}$~erg\,s$^{-1}$\,cm$^{-2}$, is in apparent disagreement with ours, although their prediction for hard X-ray flux (20--30~keV) matches the \nustar-detected flux well. A discrepancy of this magnitude may be due to the PSF of \xmm\ sampling the diffuse emission differently, resulting in different best-fit models; however, variability of the AGN cannot be excluded.

AGN variability is also suggested by the hard X-ray data, where contamination by the diffuse emission is negligible. Both the \nustar\ detection and the \suzaku/PIN upper limit put the 20--100~keV flux ($\simeq3\,\times 10^{-12}$~erg\,s$^{-1}$\,cm$^{-2}$ and $<6\,\times 10^{-12}$~erg\,s$^{-1}$\,cm$^{-2}$) below the \bepposax\ detection \citep{franceschini+2000} at $\simeq1\,\times 10^{-11}$~erg\,s$^{-1}$\,cm$^{-2}$. The \bepposax\ flux in the 20--30\,keV band, $2.6_{-1.6}^{+1.9}\,\times 10^{-12}$~erg\,s$^{-1}$\,cm$^{-2}$ \citep{piconcelli+2007} exceeds the \swiftbat\ detection limit of $\simeq1.5\,\times 10^{-12}$~erg\,s$^{-1}$\,cm$^{-2}$ \citep{vignali+2011}, as well as most extrapolations from later soft X-ray studies (e.g., $7-15\,\times 10^{-13}$~erg\,s$^{-1}$\,cm$^{-2}$ by \citealt{chiang+2013}, $6-13\,\times 10^{-13}$~erg\,s$^{-1}$\,cm$^{-2}$ by \citealt{piconcelli+2007}) and the \nustar-detected flux of $\simeq7\,\times 10^{-13}$~erg\,s$^{-1}$\,cm$^{-2}$. While it is possible that the high \bepposax\ flux was due to contamination by a nearby hard X-ray source \citep{piconcelli+2007,vignali+2011,chiang+2013}, the possibility of variability in luminosity and/or line-of-sight column density is naturally explained in our models, where the transmitted (T) component dominates the AGN spectrum.

\section{Discussion}
The IR, optical and X-ray data together form a consistent picture of the central engine in \iras. Starting with the IR data, we build this picture in \S\ref{discir} through \S\ref{discalldat}.

\subsection{The Infrared Data}\label{discir}
Our results are consistent with previous studies that mark \iras\ as AGN-dominated \citep{mrr00,sar08,mrr10,ruiz13}. By modeling the IRS spectrum together with longer wavelength data up to 1000\,$\mu$m we draw several new constraints. We clearly detect ongoing star formation in \iras\ (see also \citealt{hanhan12}). The star formation rate, at $110^{+35}_{-28}$\,M$_{\odot}$\,yr$^{-1}$, is consistent with rates seen in $z<0.2$ ULIRGs \citep{farrah03} and suggests that \iras\ is going through a significant episode of star formation despite the dominance of the AGN in the IR. The excellence of the fit is consistent with our initial assumption that there is only one current episode of star formation in \iras. Compared to the rate derived from optical observations \citep{bild08} it implies that optical data underestimate the star formation rate in \iras\ by approximately a factor of three. 

The $3\sigma$ upper limit on the age of the starburst of 50\,Myr is inconsistent with the range of 70--200\,Myr derived by \citet{pip09}. Moreover, the fit to the IR SED does not require a contribution from a second, older starburst. It is unlikely that this inconsistency arises due to model degeneracies in the IR SED fitting, since we consider all possible solutions when deducing the starburst age constraint. Instead, this implies that only the star formation seen by \citet{bild08} contributes to the IR emission, with no contribution from the event inferred by \citet{pip09}. Furthermore, since the radio jets have an age of 100--160\,Myr \citep{osullivan+2012}, it is unlikely that the ongoing star formation was triggered by the jets, or by the event that triggered the jets. This suggests that \iras\ is currently going through a {\itshape second} major epoch of luminous activity in the last 200\,Myr. This is consistent with the relatively small amount of molecular gas in this system \citep{evans98,comb11}, and suggests that \iras\ will soon become a quiescent galaxy. The upper limit on the starburst age is also consistent with the absence of Ca absorption in the optical spectrum, which suggests a relative dearth of A-type stars. 
  
We cannot, however, set useful constraints on the spatial scale of the starburst. At $z=0.442$, 1\arcsec\ corresponds to 5.7\,kpc. Compared to the spatial resolution of the IRS ($3.7$\arcsec\, and $10.5$\arcsec\, for the two low-resolution modules), and 5--10\arcsec\, for PACS, this gives a spatial resolution of 21--39\,kpc. We thus cannot say if the star formation is nuclear, spread throughout the host, or some combination of the two. 
 
Our study is the first to set IR-based constraints on the geometry of the AGN obscurer; assuming the geometry in the \citealt{efst95} models holds, then we derive $\theta^{\rm ir}_{\rm V} = 35^{+8}_{-5}\degr$ and $\theta^{\rm ir}_{\rm L} = 36^{+9}_{-6}\degr$. These values are consistent with the requirement, from the Sy2 classification, that no broad lines are visible in direct light, i.e. that $\theta^{\rm ir}_{\rm V}>\theta^{\rm ir}_{\rm L}$. Constraints on the geometry of the optical obscurer have been set, though these constraints depend on the degree of polarization and the assumed model (e.g. \citealt{bmc77}); \citet{hines99} obtain $\theta^{\rm o}_{\rm V} = 34-41\degr$ and $\theta^{\rm o}_{\rm L} = 15-33\degr$, while \citet{tran00}, who find a higher polarization, argue for $\theta^{\rm o}_{\rm V} \simeq 50\degr$ and $\theta^{\rm o}_{\rm L} \simeq 40\degr$ (see also \citealt{hines93}). Assuming that the IR and optical obscurers are co-aligned, and that $\theta^{\rm o}_{\rm V} = \theta^{\rm ir}_{\rm V}$, then our values are more consistent with those of \citet{hines99}. We find, however, that $\theta^{\rm ir}_{\rm V} \simeq \theta^{\rm ir}_{\rm L}$, whereas both \citet{hines99} and \citet{tran00} argue that $\theta^{\rm ir}_{\rm L}$ is less than $\theta^{\rm ir}_{\rm V}$, by $14\degr$ and $10\degr$, respectively. Such a difference is not entirely inconsistent with the IR-derived values, but it is also plausible that the optical obscurer has a smaller half-opening angle than the IR obscurer.

\subsection{The X-ray Data}
We start by summarizing the X-ray analysis presented in \S\,\ref{analysis_xray}. Our X-ray modeling can be separated into two branches: the simultaneous \swiftxrt\ and \nustar\ data, which feature a constant and relatively low signal-to-noise ratio across the 0.5--50~keV energy range, versus the joint \nustar\ and archival \chandra\ and \suzaku\ data, among which differences in constraining power are large and complex, and the \nustar\ contribution is smaller. The latter dataset prefers Compton-thin TD models with $N_{\rm H}\sim5\times\,10^{23}$\,cm$^{-2}$, including a tilted torus solution in which $N_{\rm H,eq}$ exceeds the Compton-thick threshold. The \swiftxrt\ and \nustar\ data lead to Compton-thick RD solutions for the AGN with each of the models, implying significantly higher intrinsic luminosity. Despite the possible issue of non-simultaneity, we consider the joint \nustar, \chandra\ and \suzaku\ dataset to be more reliable and therefore base our further discussion only on the results it provides.

We started with the T$+$R models; the \nustar\ detection disfavors the scenario where a hard, luminous and strongly absorbed T component dominates the flux above 10~keV. The preference for softer photon indices rules out the hard values ($\Gamma<1.5$) discussed in, e.g., \citet{piconcelli+2007} and \citet{chiang+2013}. Dominance of the R component in the hard X-ray band is not favored either, as the EW of Fe\,K$\alpha$ is relatively low. Instead, \iras \ resembles heavily obscured AGN in the nearby Universe, in which both T and R components contribute to the X-ray spectrum in the \nustar\ band (e.g., \citealt{puccetti+2014}, \citealt{koss+2015}, Balokovi\'{c} et al., in preparation). Both in terms of spectral components and data quality, the constraints are similar to the type~2 quasars Mrk\,34 \citep{gandhi+2014} and SDSS\,J1218$+$4706 \citep{lansbury+2015}, although both of those objects likely have higher line-of-sight column densities than \iras. The shapes and relative contributions of the T and R components depend on the geometry of the obscurer; however, the T$+$R model is only approximate, and more appropriate torus models are needed in order to derive physical constraints.

Turning to the torus models; modulo the difference in the assumed geometry and the dependence of $N_{\rm H}$ on the viewing angle, the parameters inferred from fitting the \mytorus\ and \bntorus\ models to the combined X-ray dataset are indistinguishable. Both are consistent with scenarios where the line of sight skims the edge of the torus, thus giving rise to a Sy2 classification only by a few degrees. Moreover, both imply intrinsic luminosities in the 2--10~keV band in the range 1--2$\times10^{45}$~erg\,s$^{-1}$. Notably though, the \nustar\ data are not decisive. With the \nustar\ data there is less of a $\chi^2$ gradient toward hard photon indices. However, with the assumptions used in this analysis, the same solutions can be found from the archival data alone, albeit with larger uncertainties. Relaxing the assumed spectral parameters of the plasma model for the soft X-ray part of the spectrum creates severe degeneracies such that the model becomes RD for hard $\Gamma$, i.e., opposite of the behaviour described in \S\ref{tprsection}. Although fluxes in overlapping spectral bands between \nustar, \swift, \suzaku, and \chandra\ are consistent (see \S\,\ref{sec:xray_diffuse}), spectral variability between the observations and the resulting biases in joint fitting cannot be fully excluded. Despite its coverage above 10~keV, the current \nustar\ data are insufficient to uniquely constrain the AGN spectrum, so the fits remain susceptible to the assumptions in modeling the soft X-ray data.

\subsection{The X-ray \& Infrared Data}
We now consider the X-ray and IR AGN torus models together. A cautionary note is warranted: the models for the X-ray include gas but not dust, while the models for the IR emission include only dust. In considering the two together we are thus comparing different structures. 

We first compare the derived X-ray and IR luminosities. \citet{gand09} have derived a relationship between 2-10\,keV luminosity and 12.3\,$\mu$m luminosity density for Seyferts, albeit using a sample more than two orders of magnitude less luminous than \iras, on average. Taking the $12.3\,\mu$m AGN luminosity density from Figure \ref{fig:irsed} and translating it to a predicted 2--10\,keV luminosity using the \citet{gand09} relationship yields $\sim6.3\times10^{45}$~erg\,s$^{-1}$, a factor of $\sim3$ higher than the 2--10\,keV luminosity obtained from the torus models. \citet{gand09} also see that the type~2 quasars in their sample have a lower X-ray luminosity than is predicted by their relation, and argue that the reason for this is nuclear star formation that contaminates the 12.3\,$\mu$m luminosity density. This however is an unlikely explanation for why \iras\ deviates from the relation, since the star formation in \iras\ is an order of magnitude less luminous than the AGN (the predicted $12.3\,\mu$m luminosity density of the starburst is even less than that of the AGN, but luminosities at specific wavelengths are less robust than total IR luminosities, so we are hesitant to make this comparison). This suggests that the proportionality between intrinsic X-ray and mid-IR (e.g., $\simeq12\mu$m) luminosities for AGN may flatten at high luminosities (e.g., \citealt{stern15}, but see also \citealt{asmus+2015}), or that a different relation is at work.

Turning to a comparison of the X-ray and IR geometries; it is reasonable to expect that the AGN structures producing the IR and the X-ray spectra are coaligned, which would make their respective viewing angles similar. Moreover, it is reasonable to expect their half-opening angles to be similar, motivated by comparisons of opening angles determined independently from X-ray and IR data (\citealt{brightman+2015}, \citealt{balokovic+2016}). While the high X-ray luminosity of \iras\ would make it an excellent test for the trend of decreasing torus covering factor with increasing X-ray luminosity observed for Compton-thick AGN by \citet{brightman+2015}, our modeling indicates that its line-of-sight obscuration is not significantly above the Compton-thick threshold, nor is the torus viewed close to edge-on. In this case, an independent constraint on the opening angle from the X-ray data would require a longer \nustar\ observation than the 15~ks presented here. For example, an exposure of 100~ks would provide $\simeq$10 energy bins over the 10--50~keV band with signal-to-noise ratio better than~3, sufficient to constrain the photon index within 0.1, and the torus opening angle within approximately 20\degr (quoting 90\% confidence intervals).

Coalignment of the IR and X-ray tori is consistent with the \mytorus\ and \bntorus\ results. In the \mytorus\ model, the half-opening angle, $\theta^{\rm my}_{\rm L}$ is fixed to 60$^{\circ}$, which is (just) within the 3$\sigma$ range of the IR-derived torus half-opening angle. Using \mytorus, we find that viewing angles close to 60\degr\, fit the joint X-ray dataset slightly better than edge-on ones. This is again just consistent with the result from the IR-based torus models. A useful constraint on the viewing angle can however be obtained only if the equatorial column density of the torus is assumed. For a borderline Compton-thick torus ($N_{\rm H,eq}=1\times10^{24}$~cm$^{-2}$) and the best-fit photon index ($\Gamma=1.6$), $\theta^{\rm my}_{\rm V}=\left(65\pm2\right)\degr$. If however we use $\Gamma=1.8$, then $\theta^{\rm my}_{\rm V}=\left(68_{-2}^{+4}\right)\degr$. The difference in $\chi^2$ for these two cases is negligible. This value of $\theta^{\rm my}_{\rm V}$ is still consistent with the Sy2 classification, and (within the joint error budget) with the IR-derived value, but shows that with the current data \mytorus\ constrains the geometry of the X-ray obscurer only weakly. 

The \bntorus\ constraints are stronger. Both $\theta^{\rm bn}_{\rm L}$ and $\theta^{\rm bn}_{\rm V}$ can be varied, but they cannot be independently constrained with the X-ray data. With reasonable assumptions, however, they are both consistent with the IR-based modeling results. If we fix $\theta^{\rm bn}_{\rm L}$ to 39\degr, as obtained from the IR modeling, then the best fit is found for $\theta^{\rm bn}_{\rm V}=\left(48_{-2}^{+3}\right)\degr$. This combination of $\theta^{\rm bn}_{\rm L}$ and $\theta^{\rm bn}_{\rm V}$ is within $1\sigma$ of the IR-based geometry, and represents a broad minimum in $\chi^2$ (1274, for $\nu=$1379) over the allowed range for those angles. In this case, we find $\Gamma=1.7_{-0.2}^{+0.1}$ and $N_{\rm H}=N_{\rm H,eq}=\left(4.6_{-0.9}^{+0.7}\right)\times10^{23}$~cm$^{-2}$.

Finally, the inferred bolometric luminosity from the X-ray models is consistent with that inferred from the IR models. Assuming an X-ray to bolometric correction of 50--130 \citep{marconi+2004,vasudevan+fabian-2007,lusso+2012} leads to an estimate of $L_{\rm bol}\sim(0.5-2.5)\times10^{47}$~erg\,s$^{-1}$, based on the X-ray modeling alone. Instead starting from the anisotropy-corrected IR AGN luminosity and assuming that 30\% of the bolometric emission emerges in the IR \citep{risa04} yields $\sim1.8\times10^{47}$~erg\,s$^{-1}$. Disregarding the anisotropy correction gives a still-consistent $\sim2.3\times10^{47}$~erg\,s$^{-1}$.

\begin{figure*} 
\begin{center}
\includegraphics[width=0.98\textwidth,angle=0]{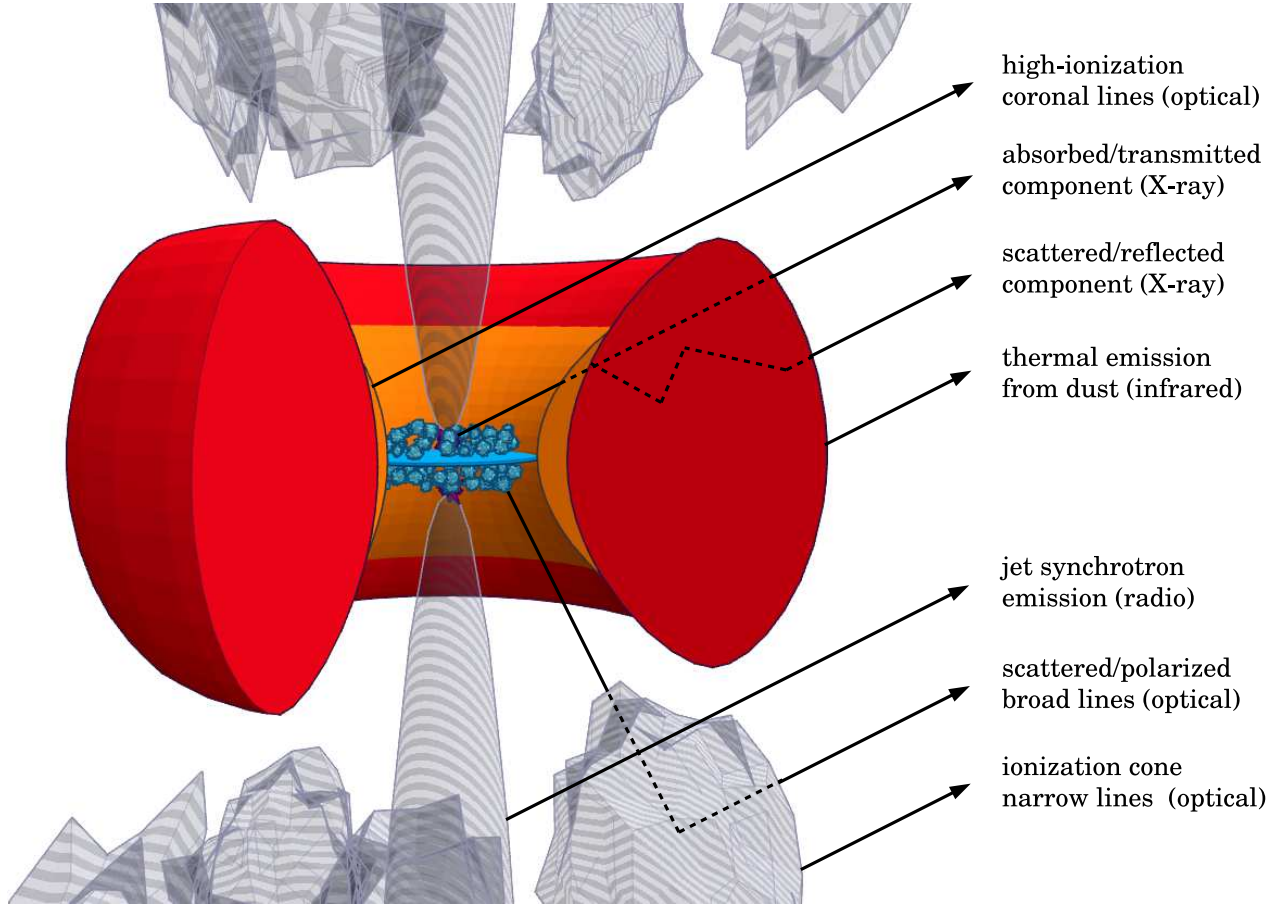}
\caption{Sketch of the \iras\ nucleus geometry that is consistent with the IR, optical and X-ray data (\S\ref{discalldat}). The observer is in the direction of the black arrows. The bulk of the torus is shown in red, and its inner wall in orange. IR emission is due to warm dust in the torus, while the coronal lines in the optical spectrum come from the inner wall of the torus. X-rays pass through the torus; we distinguish contributions from the absorbed line-of-sight component (transmission; T) and from the component due to scattering (reflection; R). The accretion disk and the broad-line clouds are shown in blue, and the jet and narrow-line clouds in the ionization cone are shown in grey. The broad line region is shielded from direct view by the vertical extent of the torus, but scattering in the ionization cones makes broad lines observable in polarized light. The ionization cones also emit narrow forbidden lines (most notably, \ion{O}{3}), and the jet is observable at radio wavelengths.}
\label{fig:torussketch}
\end{center}
\end{figure*}

\subsection{The X-ray, Infrared, \& Optical Data}\label{discalldat}
Finally, we fold in constraints from the optical data. The high excitation iron lines in Figure \ref{fig:opspec} have three possible origins; a `Coronal Line Region' (CLR) intermediate in distance between the broad and narrow line regions, the inner wall of a dusty torus, and the ISM several kpc from a `naked' Seyfert nucleus \citep{kor,pen,mur98}. The third of these possibilities predicts that [\ion{Ne}{5}] $\lambda$3426 will be $\sim$12 times stronger than [\ion{Fe}{10}] $\lambda$6375. This criterion is, at face value, consistent with our spectrum. If however we consider that [\ion{Fe}{10}] $\lambda$6375 is almost certainly contaminated by [\ion{O}{1}] $\lambda6364$, and take this contamination into account by assuming [\ion{O}{1}] $\lambda6364$/[\ion{O}{1}] $\lambda6300 = \frac{1}{3}$ then the [\ion{Ne}{5}] $\lambda$3426/[\ion{Fe}{10}] $\lambda$6375 ratio in \iras\ rises to $\sim40$. Moreover, the IR-luminous nature of \iras\ argues that an origin in a CLR and/or in a dusty torus is more plausible. 

The detection of [\ion{Fe}{10}]$\lambda$6375 but not [\ion{Fe}{14}]$\lambda$5303, if not due to differential obscuration between 5300\AA\ and 6400\AA, implies a range in hydrogen density along the line of sight of $3.0<\log n_{\rm H}\,({\rm cm}^{-3})<5.8$, and a line of sight to a distance from the central ionizing source of 0.2--20\,pc \citep{fkf}. We also note that the absence of both a 4000\AA\ break and stellar absorption features is consistent with a large population of young stars.

Combining the constraints from the IR, optical, and X-ray data is fraught with issues since the assumptions in the models were made without regard to each other. Moreover, the coronal iron lines are detected at just under $3\sigma$ significance in our spectrum. Nevertheless, the inference from the coronal iron lines of `just' seeing the inner wall of the torus is consistent with a line of sight that skims the torus -- the CLR is visible in direct light but the BLR can only be seen in scattered light \citep{tran00}. Moreover, the viewing angles inferred from the IR, optical, and X-ray data are consistent. Assuming that the geometry of the \citet{efst95} models is correct, this places the bulk of the dust column that comprises the IR-emitting torus to within a vertical height of $z=20$\,pc of the nucleus. The X-ray obscurer is thus plausibly within this distance, also. The outer `edge' of the torus is then within 125\,pc of the nucleus and the inner edge is within 2\,pc (see also \citealt{tani97}). A sketch of this geometry is shown in Figure \ref{fig:torussketch}. We do not draw detailed comparisons with literature values for AGN geometries due to the aforementioned issues with combining the datasets, but it is notable that the inner edge constraint is comparable to, though perhaps slightly larger than, that seen in AGN with similar luminosities in \citealt{burts13}.

\section{Conclusions}
We have conducted a study of \iras, an obscured hyperluminous quasar at $z=0.442$, using X-ray data from \nustar, \swift, \suzaku\ and \chandra, infrared data from {\itshape Spitzer} and {\itshape Herschel}, and an optical spectrum from Palomar. We apply radiative transfer models to the infrared data to measure rates of ongoing star formation in the host galaxy, and to constrain the properties of the infrared obscurer around the AGN. We apply two types of models to the X-ray data -- a T$+$R (phenomenological) model and the \mytorus/\bntorus\ (geometrical) models -- to constrain the properties of the X-ray obscurer. We then fold in a distance constraint from the optical spectrum to construct a picture of the geometry of the structure around the AGN in this archetype object. Our conclusions are:

1 - The infrared data can be reproduced by a combination of an AGN and a starburst. The total infrared (rest-frame 1-1000$\mu$m) luminosity is $6.76\pm0.20\times10^{46}$~erg\,s$^{-1}$, with a contribution from the AGN of $5.94^{+0.26}_{-0.27}\times10^{46}$~erg\,s$^{-1}$. The starburst is required in the fit at $3.7\sigma$ confidence, with a luminosity of $5.54\pm1.48\times10^{45}$~erg\,s$^{-1}$, corresponding to a star formation rate of $(110^{+35}_{-28})$\,M$_{\odot}$ yr$^{-1}$. Accounting for the anisotropic emission in the AGN models leads to an intrinsic AGN infrared luminosity of $\sim4.9\times10^{46}$~erg\,s$^{-1}$ and a total infrared luminosity (assuming the starburst emission is isotropic) of $\sim5.5\times10^{46}$~erg\,s$^{-1}$. The ratio between the mid-infrared and 2-10~keV luminosities may deviate from that seen in lower luminosity Seyferts, consistent with an intrinsically different relation at very high luminosities. 

2 - The infrared AGN torus model has a viewing angle (from pole-on) of $\theta^{\rm ir}_{\rm V} = \left(35^{+8}_{-5}\right)\degr$ and a half-opening angle of $\theta^{\rm ir}_{\rm L} = \left(36^{+9}_{-6}\right)\degr$. The starburst model is consistent with an age for the starburst of $<50Myr$. The AGN model parameters are consistent with the requirement, from the Sy2 classification, that no broad lines are visible in direct light, i.e. that $\theta^{\rm ir}_{\rm V}>\theta^{\rm ir}_{\rm L}$. They are also consistent with the geometry of the (assumed) biconical structure giving rise to the optical emission lines. The star formation rate is comparable to those seen in lower redshift ULIRGs, and suggests that the host of \iras\ is going through a significant stellar mass assembly event. The age constraint is however inconsistent with both the age of the radio jets (120--160\,Myr) and the age of a previous starburst event (70--200\,Myr). This suggests that \iras\ underwent at least two epochs of luminous activity in the last $\sim200$\,Myr: one approximately $150$\,Myr ago, and one ongoing. 

3 - The X-ray model fits are consistent with $\Gamma\simeq1.8$ and $N_{\rm H}\sim5\times10^{23}$~cm$^{-2}$ (T$+$R: $\Gamma=1.8_{-0.4}^{+0.2}$ and $N_{\rm H}=\left(5_{-2}^{+3}\right)\times10^{23}$~cm$^{-2}$; torus models: $\Gamma=1.7_{-0.2}^{+0.1}$ and $N_{\rm H}=N_{\rm H,eq}=\left(4.6_{-0.9}^{+0.7}\right)\times10^{23}$~cm$^{-2}$). The soft X-ray data alone drive the fit toward hard photon indices ($\Gamma<1.5$) and a TD solution, but the addition of \nustar\ data results in a solution where T and R components contribute at comparable levels, and rules out an RD scenario in which line-of-sight obscuration and intrinsic luminosity would be much higher.

4 - The constraints on the AGN obscurer geometry from the X-ray data are, with reasonable assumptions, consistent with those inferred from the infrared data. Fixing $\theta^{\rm bn}_{\rm L}$ to 39$^{\circ}$ in the \bntorus\ model gives a best-fit viewing angle of $\theta^{\rm bn}_{\rm V}=\left(48_{-2}^{+3}\right)^{\circ}$. This combination of $\theta^{\rm bn}_{\rm L}$ and $\theta^{\rm bn}_{\rm V}$ coincides with a broad minimum in $\chi^2$ for all the \bntorus\ model fits, and is within $1\sigma$ of the IR-based half-opening and viewing angles. The \mytorus\ model constraints are similar, though weaker. The X-ray and infrared torus models are thus both consistent with scenarios where the line of sight viewing angle is close to the half-opening angle. The data do not favor extreme geometries, such as edge-on viewing angle, or tori that are disk-like ($\theta^{\rm bn}_{\rm L}\rightarrow90^{\circ}$) or sphere-like ($\theta^{\rm bn}_{\rm L}\rightarrow0^{\circ}$). This `skimming' of the edge of the torus by the line-of-sight viewing angle suggests that, had \iras\ been viewed at a viewing angle smaller by only a few degrees, it would have been classified as a broad-line object in direct light. 

5 - The constraints on the bolometric luminosity of \iras\ from the X-ray and infrared data are also consistent with each other. The intrinsic 2--10~keV luminosity lies in the range 1--2$\times10^{45}$~erg\,s$^{-1}$. Assuming a bolometric correction of 50--130 leads to an estimate of $L_{\rm bol}\sim(0.5-2.5)\times10^{47}$~erg\,s$^{-1}$. Instead starting from the intrinsic AGN luminosity derived from IR modeling, and assuming that 30\% of the bolometric emission emerges in the infrared, yields $\sim1.8\times10^{47}$~erg\,s$^{-1}$.

6 - The detection of high excitation iron lines in the optical spectrum provides further constraints on the geometry of the AGN obscurer. If these lines arise in a Coronal Line Region (CLR) then their detection is consistent with a line of sight that skims the torus - the CLR is visible in direct light but the BLR can only be seen in scattered light. Taking the distance constraints from the detection of $[$\ion{Fe}{10}$]\lambda$6374 but not $[$\ion{Fe}{14}$]\lambda$5303 then places the bulk of the dust column that comprises the IR-emitting torus to within a vertical height of 20\,pc of the nucleus. The X-ray obscurer is thus plausibly within this distance, also. Assuming that the geometry of the infrared model is correct then places the outer `edge' of the IR-emitting torus within 125\,pc of the nucleus and the inner edge within 2\,pc. These values have large systematic uncertainties that are difficult to estimate, and are based on the aforementioned combining of assumptions across disparate models.

7 - The joint X-ray dataset, despite its broadband coverage, is insufficient to provide constraints on the AGN torus geometry without keeping some model parameters fixed, and/or without constraints from the infrared and optical data. The 15-ks \nustar\ observation, despite the $\simeq13\,\sigma$ detection above 10\,keV, does not constrain the AGN spectrum of \iras\ substantially better than the archival data below 10\,keV. The joint X-ray dataset gives less of a $\chi^2$ gradient toward hard photon indices, however, similar solutions can be found from the archival data alone, albeit with larger uncertainties. Moreover, both $\theta^{\rm bn}_{\rm L}$ and $\theta^{\rm bn}_{\rm V}$ cannot be independently constrained. Despite its coverage above 10~keV, the current \nustar\ data are not of sufficient quality to uniquely constrain the AGN spectrum, so the fits remain susceptible to assumptions. A longer \nustar\ observation of \iras\ is essential for constraining the structure of the torus directly from the X-ray band.

\acknowledgments
We thank the referee for a very helpful report. This work was supported under NASA Contract No.~NNG08FD60C, and made use of data from the \nustar \ mission, a project led by the California Institute of Technology, managed by the Jet Propulsion Laboratory, and funded by the National Aeronautics and Space Administration. We thank the \nustar \ Operations, Software and Calibration teams for support with the execution and analysis of these observations. This research has made use of the \nustar \ Data Analysis Software (NuSTARDAS) jointly developed by the ASI Science Data Center (ASDC, Italy) and the California Institute of Technology (USA). Herschel is an ESA space observatory with science instruments provided by European-led Principal Investigator consortia and with important participation from NASA. PACS has been developed by a consortium of institutes led by MPE (Germany) and including UVIE (Austria); KU Leuven, CSL, IMEC (Belgium); CEA, LAM (France); MPIA (Germany); INAF-IFSI/OAA/OAP/OAT, LENS, SISSA (Italy); IAC (Spain). This development has been supported by the funding agencies BMVIT (Austria), ESA-PRODEX (Belgium), CEA/CNES (France), DLR (Germany), ASI/INAF (Italy), and CICYT/MCYT (Spain). This work is based in part on observations made with the Spitzer Space Telescope, which is operated by the Jet Propulsion Laboratory, California Institute of Technology under a contract with NASA. Part of this work is based on archival data, software and online services provided by the ASDC. This research has made use of NASA's Astrophysics Data System. We acknowledge support from the NASA Earth and Space Science Fellowship Program grant NNX14AQ07H (MB), CONICYT-Chile grants Basal-CATA PFB-06/2007 (FEB, CR), FONDECYT Regular 1141218 (FEB, CR), "EMBIGGEN" Anillo ACT1101 (FEB, CR), and the Ministry of Economy, Development, and Tourism's Millennium Science Initiative through grant IC120009, awarded to The Millennium Institute of Astrophysics, MAS (FEB). ACF acknowledges ERC Advanced Grant Feedback 340442.

{\it Facilities:} \facility{NuSTAR}, \facility{Swift}, \facility{Suzaku}, \facility{Chandra}, \facility{Spitzer}, \facility{Herschel}, \facility{Palomar}.

\end{document}